\documentclass[aps, superscriptaddress, prd, 10pt, twocolumn, floatfix ]{revtex4}
\usepackage{amsmath, amssymb, amsfonts}
\usepackage{hyperref}
\usepackage{graphicx}
\usepackage{color}

\newcommand{\figref}[1]{Fig:~\ref{#1}}

\begin{document}

\title{Charm Quark Energy Loss In Infinite QCD Matter Using a Parton Cascade Model}

\author{Mohammed Younus} \email{younus.presi@gmail.com}
\affiliation{Variable Energy Cyclotron Centre, 1/AF Bidhan Nagar
Road, Kolkata 700064, INDIA}
\author{Christopher E. Coleman-Smith}
\affiliation{Duke University, Dept. of Physics, 139 Science Drive,
Box 90305, Durham NC 27708, U. S. A.}
\author{Steffen A. Bass} \email{bass@phy.duke.edu}
\affiliation{Duke University, Dept. of Physics, 139 Science Drive,
Box 90305, Durham NC 27708, U. S. A.}
\author{Dinesh K. Srivastava} \email{dinesh@vecc.gov.in}
\affiliation{Variable Energy Cyclotron Centre, 1/AF Bidhan Nagar
Road, Kolkata 700064, INDIA}

\begin{abstract}
  We utilize the Parton Cascade Model to study the evolution of charm
  quarks propagating through a thermal brick of QCD matter.  We  determine the
  energy loss and the transport coefficient '$\hat{q}$' for charm quarks. The calculations are done at a
  constant temperature of $350$~MeV and the results are compared to  analytical
  calculations of heavy quark energy loss in order to validate the applicability of using a Parton Cascade
  Model for the study of heavy quarks dynamics in hot and dense QCD matter.
\end{abstract}

\maketitle

\section{Introduction}
\label{secn-A}

Relativistic heavy ion collision at RHIC and the LHC have given rise to a new phase of matter. When two heavy
ions collide, a system of deconfined gluons and quarks within a very small volume is created.  The initial energy density within this volume is found to be on the order of 30 GeV/fm$^3$.  This state of matter as we know today is called quark gluon plasma~\cite{qgp1,qgp2}.  The
study of QGP is particularly important as it aims to produce a condition which resembles the period when
universe was only a few microseconds old.  However, since this exotic system created in the experiments exists only for a very short period of time and is not directly observable, only signals
originating from the matter itself that survive and are measured after the collisions can provide
a window into the
nature of the QGP~\cite{qgprev1,qgprev2}.

One of the prominent signatures coming out of the QGP phase is jet quenching: High momentum hadron spectra are observed to be highly
suppressed relative to those in proton on proton collisions~\cite{suppress1,suppress2}, suggesting a quenching effect due to deconfined matter. A similar effect is observed for high $p_T$ charm or beauty quarks with most recent results showing suppression of D or
B mesons to same order as that of light partons~\cite{exp1}.
Calculations from hydrodynamics also give a rough estimate of the ratio of thermalization time for heavy quarks and
light partons~\cite{therm1}, $\frac{\tau_{Q}} {\tau_{q/g}}\sim\frac{M_Q}{T}$. For $M_Q$ = 1.35--4.5 GeV and T = 300
MeV, this ratio is found to be$\sim$ 5 and suggests that relaxation time for heavy quarks
is larger than that of light quarks and gluons.
If thermalization time, $\tau_{q/g}$, is taken to be $\mathcal{O}$(1fm/c), and
if equilibrium temperature, $T_i$ and freeze-out temperature, $T_f$ are taken as 300 MeV and 170 MeV~\cite{therm2} respectively, then the lifetime of the QGP can be
approximately shown to be 5 fm/c. This might imply that the heavy quark relaxation time for T= 300 MeV is comparable to  the QGP lifetime at this condition.
Even if the heavy quark is subjected to large suppression~\cite{exp1}, it may not
fully thermalize in the QGP. Overall, the study of heavy quark dynamics is slowly emerging as one of
the most active fields of research in heavy ion collision physics.

Various theoretical calculations and phenomenological models of heavy quarks energy loss have appeared in
recent years~\cite{therm3}--\cite{theo5}.  Elastic scattering and inelastic gluon emission are the
major mechanisms by which a heavy quark may lose energy in the presence of a thermal medium.  In most of these
earlier works, collisional energy loss seems to dominate in the lower momentum region while radiative energy
loss emerges as the chief mechanism for higher momenta charms.

Transport models attempt to fully describe the dynamics of the time-evolution of a heavy ion collision. The Parton
Cascade Model is one such model~\cite{pcm1,pcmother1,pcmother2}. It is based on the Boltzmann Equation
 and does not include any equilibration assumptions. However the calculations must be well calibrated and
validated under controlled conditions before utilizing them for meaningful predictions. Performing
this validation for the medium evolution of heavy quarks is the purpose of this work.

\section{Parton Cascade Model}
\label{secn-B}
The Parton Cascade Model VNI/BMS~\cite{pcmearly, pcm2, pcm3} forms the basis for our present study.  This
model can be used to study the full time evolution of hard probes in a thermal QCD medium. The PCM has been
used to study gluons and lighter quarks as hard probes of the QGP. In the current work we use VNI/BMS to study the evolution of
charm quarks in an infinite QGP medium for the first time.  The purpose of this study is to provide
a verifiable benchmark calculation to validate the model and subsequently apply it to the more
complex and dynamic regime of a heavy-ion collision.

The infinite QGP medium is modeled by taking a box of finite volume  with periodic boundary
conditions. This provides a system of infinite matter at fixed temperature.  The matter inside the box
consists of
thermalized quarks and gluons (QGP) which are being generated using thermal distributions at a given
temperature and zero chemical potential.  We insert a charm quark with the four momentum $p^{\mu}
=\{0,0,p_{z},E=\sqrt{p_{z}^{2}+M_{c}^{2}}\}$, into the box and let it evolve according to the Relativistic
Boltzmann Equation given by,
\begin{equation}
p^{\mu}\frac{\partial F_{k}(x,\vec{p})}{\partial x^{\mu}}=\,\sum_{processes:i} C_i [F],
\end{equation}
where $F_{k}(x,\vec{p})$ is the  single particle phase space distribution and the collision term on
r.h.s. is a non-linear functional of phase space distribution terms inside an integral.

We have included the matrix elements for all $2\rightarrow2$ binary elastic scattering processes for charm
interaction with gluons or light quarks($u,d,s$) and $2\rightarrow n$ process for radiative (brehmsstrahlung)
corrections after each scattering.

\subsection{Elastic scattering of charm quark}
The elastic processes included are
\begin{align}
  cg&\rightarrow cg, \\
  cq(\bar{q})&\rightarrow cq(\bar{q}).\nonumber
\end{align}
The corresponding differential scattering cross section is defined to be,
\begin{equation}
 \frac{d\hat{\sigma}}{dQ^2}=\frac{1}{16\pi(\hat{s}-M_{c}^2)^2}\sum{|\mathcal{M}|^2}.
\end{equation}
The total cross section is also calculated and used in the calculations to select interacting pairs.  The
total cross section can be shown to be,
\begin{equation}
\hat{\sigma}_{tot}=\sum_{c,d}\int_{p_{Tmin}^2}^{\hat{s}}
\left(\frac{d\hat{\sigma}}{dQ^2}\right)_{ab\rightarrow cd}dQ^2.
\end{equation}
The invariant transition amplitudes, $|\mathcal{M}|^2$ for elastic scattering which can be calculated or obtained
from~\cite{invM}, are shown below for $q(\bar{q})c\rightarrow q(\bar{q})c$ ,
\begin{align}
\sum{|\mathcal{M}|^2}=\frac{64\pi^2\alpha_{s}^2}{9}\frac{(M_{c}^2-\hat{u})^2+
(\hat{s}-M_{c}^2)^2+2M_{c}^2\hat{t}}{(\hat{t}-\mu_{D}^2)^2}.
\end{align}
While, for $gc\rightarrow gc$,
\begin{equation}
\sum{|\mathcal{M}|^2}=\pi^2\alpha_{s}^2[g1+g2+g3+g4+g5+g6]\,,\nonumber
\end{equation}

\begin{align}
g1&=32\frac{(\hat{s}-M_{c}^2)(M_{c}^2-\hat{u})}{(\hat{t}-\mu_{D}^2)^2}\,,\nonumber\\
g2&=\frac{64}{9}\frac{(\hat{s}-M_{c}^2)(M_{c}^2-\hat{u})+2M_{c}^2(\hat{s}+M_{c}^2)}
{(\hat{s}-M_{c}^2)^2}\,,\nonumber\\
g3&=\frac{64}{9}\frac{(\hat{s}-M_{c}^2)(M_{c}^2-\hat{u})
+2M_{c}^2(M_{c}^2+\hat{u})}{(M_{c}^2-\hat{u})^2}\,,\nonumber\\
g4&=\frac{16}{9}\frac{M_{c}^2(4M_{c}^2-\hat{t})}{(\hat{s}-M_{c}^2)(M_{c}^2-\hat{u})}\,,\nonumber\\
g5&=16\frac{(\hat{s}-M_{c}^2)(M_{c}^2-\hat{u})+M_{c}^2(\hat{s}-\hat{u})}
{(\hat{t}-\mu_{D}^2)(\hat{s}-M_{c}^2)}\,,\nonumber\\
g6&=-16\frac{(\hat{s}-M_{c}^2)(M_{c}^2-\hat{u})-M_{c}^2(\hat{s}-\hat{u})}
{(\hat{t}-\mu_{D}^2)(M_{c}^2-\hat{u})} .
\end{align}

In order to regularize the cross sections we have used the thermal mass of QGP medium which is defined as
$\mu_D$ = $\sqrt{(2N_c+N_f)/6}gT$, where $g=\sqrt{4\pi\alpha_s}$ and $\alpha_s$ is the strong coupling
constant. $N_f$, no. of flavours and $N_c$, no. of colours are taken 4 and 3 respectively. We have kept
$\alpha_s$=0.3 fixed for the entire calculation. While the PCM has the capability of using a running
or temperature-dependent coupling constant, keeping it a fixed value allows us to easily compare
our calculations to analytic expressions for the same quantities, which is the main purpose of
the present work.

The Boltzmann transport equation is then solved numerically via Monte Carlo algorithms, a geometric
interpretation of the cross section is used to select which collisions will occur.

\subsection{Charm Quark Radiation }
It is known that collisional loss alone is unable to explain the data showing suppression of
D mesons at LHC~\cite{collrad}. On the one hand, the hard thermal loop(HTL)
approximation~\cite{HTL1,HTL2} predicts a large drag on heavy quarks which is much bigger than what
experimental data has suggested, while the radiative corrections to heavy quark energy loss when combined with
elastic scattering are able to explain the results agreeably~\cite{collrad}.

In our calculations, radiative corrections are included in form of time-like branching of the probe charm into
a final charm and a shower of radiated partons using Altarelli-Parisi(AP) splitting function~\cite{ap}. The basic idea is that during a binary scattering the outgoing
partons may acquire some virtuality. These partons are allowed to radiate a shower of partons until their
virtuality decreases to some preassigned cutoff value, $\mu_0^2$ ($\approx M_{c}^2$ for charm quarks).

Any quark subjected to multiple collision may radiate a shower of partons as has been discussed earlier~\cite{BetheHeitler,LPM2,deadcone1}.  However emission of
multiple partons within a certain length scale may lead to a reduction of the bremsstrahlung cross-sections
which we can briefly discuss here.  This reduction in emitted gluon spectrum is known as Landau Pomeranchuk
Migdal (LPM) effect~\cite{LPM1}. This arises from the fact that if the formation time of an emitted gluon
after a $Qq(Qg)$ scattering is larger than the typical mean free path of the heavy quark itself, then a
gluon emitted from the next scattering centre may interact coherently with the initial gluon. This
interference of emitted gluons may continue if there are a number of scattering centres before the shower of gluons
dissociates itself completely from the emitting parton.
Radiative energy loss via the LPM effect has previously been calculated  for heavy quarks by~\cite{LPM3, deadcone2}.
The LPM effect in radiative corrections to charm quark energy loss has been utilized to describe the observed
suppression of single non-photonic electrons~\cite{theo3}.

In the PCM, the LPM effect has been implemented using a MC algorithm \cite{pcm4} first proposed by
Zapp and Wiedemann \cite{pcm5}:
This method is particularly appealing since it requires no artificial parametrization of the radiative process, it is a  purely probabilistic medium induced modification.

After the production of a parton shower via an inelastic collision, the hardest radiated gluon is selected to represent the shower as the probe and re-interact with the medium. This reflects the dominance of the gluon rescattering in the interference process. The formation times
\begin{equation}
\label{eqn-formation-time}
\tau_f^{0} = \sum_{branchings}\frac{\omega}{\mathbf{k}_{\perp}^2},
\end{equation}
 for each branching during the parton shower leading up to the production of the probe gluon are summed. The heavy quark is allowed to propagate through the medium and rescatter elastically during this time, the remainder of the partons from the radiation event propagate spatially but may not interact. Each time the probe gluon rescatters its formation time is recalculated as
 \begin{equation}
   \label{eqn-formation-time-recalc}
   \tau_f^{n} = \frac{\omega}{\left(\mathbf{k}_{\perp} + \sum_{i=1}^{n} \mathbf{q}_{\perp,i}\right)^2}.
 \end{equation}
this simulates the emission of the shower from $n$ centers which transfer their momentum coherently. After this formation time expires the radiation is considered to have separated from the initiating heavy quark and all partons may once again interact and radiate.

\begin{figure}[tb]
\includegraphics[height=3.0in,width=2.9in,angle=270]{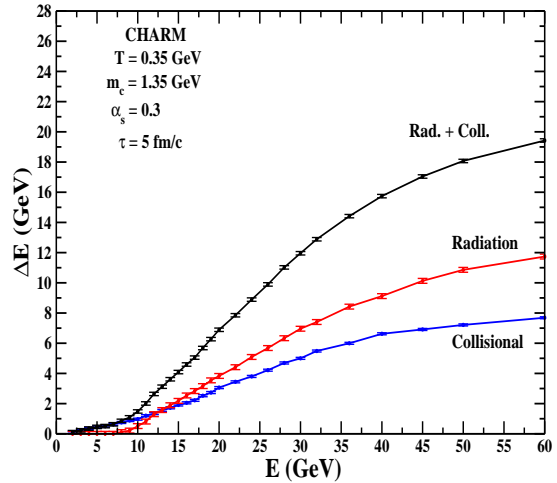}
\caption{(Color online)Energy loss for different initial charm 
energies after $5$~fm/c of propagation.}
\label{fig1}
\end{figure}

\begin{figure}[tb]
\includegraphics[height=3.0in,width=2.9in,angle=270]{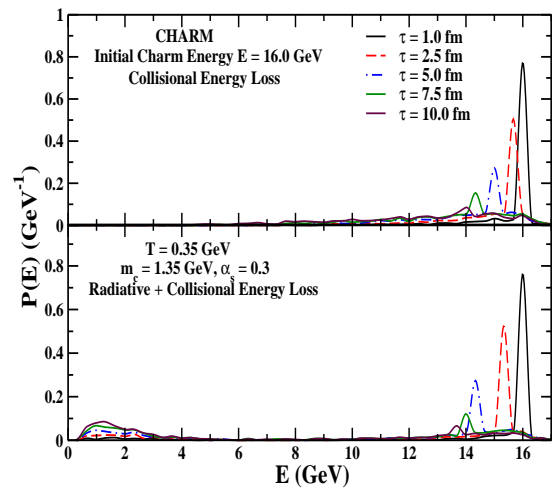}
\caption{(Color online)Energy profile for 16 GeV charm for different time}
\label{fig2}
\end{figure}

\begin{figure}[tb]
\includegraphics[height=3.0in,width=2.9in,angle=270]{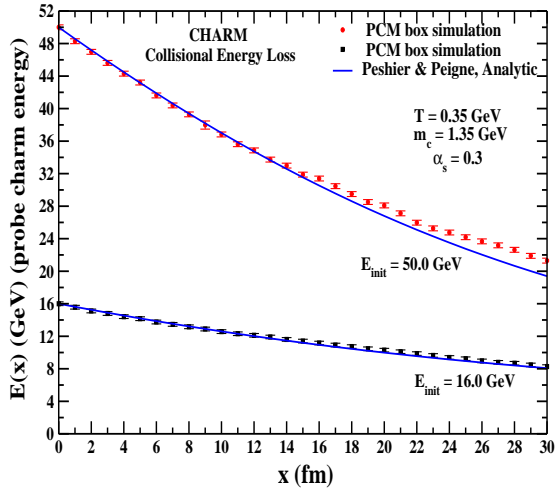}
\caption{(Color online)Energy of probe charm with distance traveled for
elastic scattering only}
\label{fig3}
\end{figure}

\begin{figure}[htb]
\includegraphics[height=3.0in,width=2.91in,angle=270]{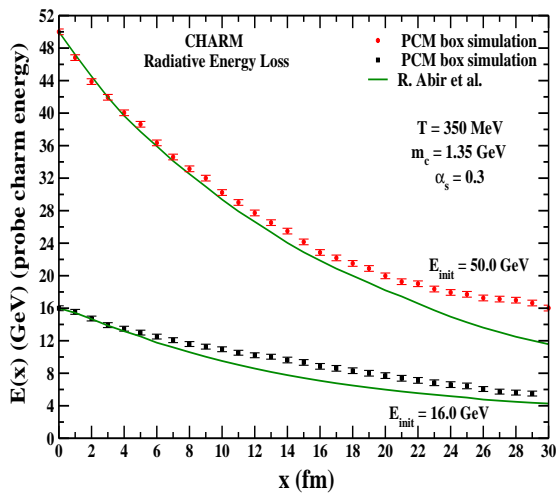}
\caption{(Color online)Energy of probe charm with distance traveled for
radiative energy loss only}
\label{fig4}
\end{figure}

\section{Results and Discussion}
\label{secn-C}

In our calculations we have set the strong coupling constant to a fixed value of $\alpha_s=0.3$
to allow comparison with analytical calculations and
other transport models. The temperature is set to T=350~MeV, 
which is roughly the average temperature of the QGP phase attained at RHIC
energies. The mass of charm is taken as $M_c =1.35$~GeV. For 
future applications of our model to nucleus-nucleus collisions we shall use a running coupling constant instead.

In \figref{fig1}, we show energy loss,'$\Delta E$', of charm quark over a given path length($L\sim5 fm$) as a 
function of its initial energy.
For discussions on path length dependence of energy loss evolution, other figures in this paper will be referred next.

We may now return to \figref{fig1} for detailed discussions. The loss due to elastic scattering, radiation and 
total loss are
shown separately in the same figure. We may observe that collisional loss dominates over radiation up to $7-12$~
GeV of initial charm energy, and beyond this energy regime, charm quark radiation dominates and 
contributes more towards the energy loss. 
However the radiation too appears to decrease for very high energy charms which may be 
the factor behind the decrease in charm quark suppression in high momentum regions. 
With increasing initial energy of the charm quark, the rise in energy loss(both 
collisional and radiative) over the 5 $fm$ length 
appears to level off -- we can see this in particular prominent for elastic energy loss. 
We feel that as momentum of charm increases, the average no. of elastic scattering tends to saturate so
that the collisional energy loss tends to saturate. But let us recall that in our case, radiation takes
place only after elastic scattering, and as the no. of scattering saturates ultimately, so
does the radiative loss for very high energy charm quarks. This particular trend may be 
reflected in the decrease of nuclear suppression factor, $R_{AA}$, of heavy quark in the final 
heavy meson spectra at high momentum region~\cite{exp1}.
Our findings are consistent with e.g.~\cite{therm1}, where it was discussed 
that for small coupling, $\alpha_s$,
collisional loss tends to dominate for low and intermediate energy charm(for 
$\gamma v_Q\sim 1\,,\gamma=(1-\beta^2)^{-1/2}$) while for higher energetic heavy quarks we have bremsstrahlung
(for $\gamma v_Q\sim 1/g\,,g=\sqrt{4\pi\alpha_s}$) dominating over 
collisional energy-loss. Other discussions on the topic are given in~\cite{book1}.

In \figref{fig2} we show the energy profile of a $16$~GeV charm  after several time intervals of
propagation through the thermal medium. Here $P(E)$ can be defined as $=\frac{1}{N}\frac{dN}{dE}$.  The energy loss due to collisional and collisional+radiative  processes is shown
separately in the same figure. The collisional loss (upper panel) shows a shift in the position of the peak  with long tail like
structure extending towards the low energy regions. A recent study of charm quark energy
profile using a Langevin equation along with a hydrodynamical background has instead shown a more Gaussian like distribution
~\cite{shangauss}. Some other discussions on the differences between Boltzmann and Langevin equations for heavy quark dynamics are also given in~\cite{shangauss}. Additionally we find that inclusion of radiative corrections
brings about a significant change in the profile and indicates that for high energy charm quarks the effect of radiative loss
is much greater than collisional loss, with the bulk of of 16.0 GeV charm quarks ultimately shifting to very low
energy($< 2.0$~GeV) regions after $10$~fm.

Next we study the evolution of charm quark energy as a function of distance traveled
through the medium in \figref{fig3} and \figref{fig4}. The calculation uses two
different initial energies ($16$~GeV and $50$~GeV respectively) for charm.
Collisional loss and radiative loss are shown in these two figures separately -- the radiative
energy-loss figure was obtained by subtracting the elastic energy-loss calculation from the full calculation that included
collisional+radiative energy-loss. We would like to elucidate the fact that these two diagrams show energy
of charm quark after each '$fm$' of path length traversed and shows the path length behaviour of charm
quark. These plots may be used to give the total energy loss of charm for comparison to \figref{fig1}.

Now let us discuss Figs.~\figref{fig3} and \figref{fig2} in detail.
The curves for the 50~GeV charm quarks show a clear distinction between the radiative and
collisional energy-loss mechanisms: whereas the collisional energy-loss shows initially a linear
behavior, the radiative energy-loss leads to a much stronger, near quadratic, fall-off in the
energy for the first 20~fm/c. For the charm quarks with an initial energy of 16~GeV the differences
are far less pronounced, but even here a ratio between the two curves would yield interesting
differences.
For both cases, we compare our results to analytical calculations of
$dE/dx$. For collisional loss we have used an analytical form
by Peshier and
Peigne~\cite{coll2} which can be written as:
\begin{align}
\frac{dE}{dx}=\frac{4\pi\alpha_{s}^2T^2}{3}\left[\left(1+
\frac{N_f}{6}\right).\right.\nonumber\\
\left.\ln{\frac{E_p(x)T}{\mu_{D}^2}}+\frac{2}{9}
\ln{\frac{E_p(x)T}{M_{c}^2}}+c(N_f)\right]\nonumber\\
\end{align}

Both for our PCM calculation as well as for the analytical expression we have used the following values for the parameters in order to compare the two. They are:
a medium temperature of T= 350 MeV (applicable for RHIC-QGP system), a charm mass of $M_c$=1.35 GeV,
no. of flavours and colours, $N_f$=4, $N_c$=3, a fixed coupling strength of $\alpha_s=0.3$, and a screening mass $\mu_D=\sqrt{(2N_c+N_f)/6}gT$. We find that for the above set of parameters, the
PCM results show good agreement to the predictions from the analytical expression, validating
our computational setup and approach.

Next we move over to results on charm quark radiative energy loss~\figref{fig4}.
The radiative energy loss  is compared to an analytical calculation by R. Abir et al~\cite{rad1} shown
below:

\begin{eqnarray}
\frac{dE}{dx}&=&24\alpha^3_s\left(\rho_q+\frac{9}{4}\rho_g\right)\frac{1}{\mu_g}(1-\beta_1)\nonumber\\
             &\times&\left(\frac{1}{\sqrt{(1-\beta_1)}}[\log{(\beta_1)}^{-1}]^{1/2}-1\right)\mathcal{F}(\delta)\nonumber
\end{eqnarray}
where
\begin{align}
\mathcal{F}(\delta)=2\delta-\frac{1}{2}\log{\left(\frac{1+M^2_ce^{2\delta}/s}{1+M^2_ce^{-2\delta}/s}\right)}\nonumber\\
-\frac{M^2_c\cosh{\delta}/s}{1+2M^2_c\cosh{\delta}/s+M^4_c/s^2}\,,\nonumber\\
\delta=\frac{1}{2}\log\left[\frac{\log{\beta_1^{-1}}}{(1-\beta_1)}\left(1+\sqrt{1-\frac{(1-\beta_1)^{1/2}}{[\log\beta_1^{-1}]^{1/2}}}\right)^2\right]\,,\nonumber\\
s=E^2(1+\beta_0)^2\,,\,\beta_1=\frac{g^2}{C}\frac{T}{E}\,,\,\beta_0=(1-M^2_c/E^2)^{1/2}\,,\nonumber\\
C=\frac{3}{2}-\frac{M^2_c}{4ET}+\frac{M^4_c}{48E^2T^2\beta_0}\log\left[\frac{M^2_c+6ET(1+\beta_0)}{M^2_c+6ET(1-\beta_0)}\right]\nonumber\\
\end{align}

As in the elastic energy-loss case, we have used identical values for parameters in the PCM calculation and in the analytic case, such as $T=$350 MeV, $M_c$= 1.35 GeV, $\alpha_s$= 0.3, $N_f$= 4, $N_c$= 3 and $\mu_D=\sqrt{(2N_c+N_f)/6}gT$.

Note, however, that the calculations of~\cite{rad1} is carried out in the  Bethe-Heitler limit of
radiative energy loss with the effects of the dead-cone formalism being explicitly included in the calculation. The authors of \cite{rad1} state that the
LPM effect if added would only affect a marginal change in the final gluon emission spectrum which is clearly not what our results suggest.
The PCM simulation explicitly takes the LPM effect into account as discussed in the previous sections. We do find that our simulation results for coherent gluon emission of
charm quarks agrees reasonably well to that of the analytical calculation upto $x\,=$ 5--6~fm, supporting the claim that modifications to the heavy-quark emission spectrum due to the LPM effect for this particular medium length, are modest. For $x\,>$ 6 fm, however, the simulation result involving LPM effect and analytical curve in the BH limit move apart from each other. When we change the energy of charm probe,
$E_c$, from 16 GeV to 50 GeV, the differences between BH and LPM radiative mechanisms 
increase and become more profound and visible. This may be indicative of the 
rising importance of the coherent gluon emission effects at higher charm quark energies. We would also like to 
highlight our choice of parameter, $N_f$= 4, which introduces an 10$\%$ uncertainty in 
the total interaction cross-section assuming that the medium partons in our calculations 
are massless.

Overall we are confident that the comparison and agreement between PCM and the analytical calculations validates the PCM approach to heavy-quark
energy loss and allows us to utilize the PCM for observables and calculations that are beyond the
scope of analytical approaches, e.g. in the rapidly evolving non-equilibrium domain of
ultra-relativistic heavy-ion collisions.

Next let us move over to our calculation of transverse momentum broadening per unit length of
charm quarks also known as the transport coefficient $\hat{q}$~\cite{coeff1,coeff2}. In other words $\langle\hat{q}\rangle$ is a jet-quenching parameter calculated as a measure of momentum broadening
within various energy loss models. Also the term 'transverse' refers to the direction perpendicular to the original direction of propagation and consequently for a jet of partons in the medium, the average or mean momentum of the jet remains unchanged while the momenta of each parton show broadening resulting in the redistribution of the transverse momentum spectrum of the jet partons. Some
recent calculations have suggested values of this coefficient ranging from 0.5--20 GeV$^2$/fm~\cite{coeff3}
for light quarks. For heavy quarks, it was calculated in~\cite{coeff4} which showed the value of $\hat{q}\sim$
0.3--0.7 GeV$^2$/fm. More detailed discussions and recent results on $\hat{q}$ of partons and heavy quarks can be found in~\cite{coeff5,coeff6}.

Generally, the transport coefficient $\hat{q}$ can be defined as:
\begin{equation}
\frac{d(\Delta p_{T}^2)}{dx}=\hat{q}=\rho
\int d^2q_\bot q_{\bot}^2
\frac{d\sigma}{d^2q_\bot}
\end{equation}
where $\frac{d\sigma}{d^2q_\bot}$ is the differential scattering cross-section of $Q$ with medium quarks and
gluons.  In case of a Monte Carlo simulation this definition can be rewritten as:
\begin{equation}
 \hat{q}=\frac{1}{l_x}\sum_{i=1}^{N_{coll}}(\Delta p_{T,i})^2
\end{equation}

For $T= 350$~MeV and the probe charm energy of 16~GeV, we find
$\hat{q}$ to be $1.2$~GeV$^2$/fm with an uncertainty of $\pm0.2$~GeV$^2$/fm, while for
charm energy of 50~GeV $\hat{q}$ is calculated to be 1.1 GeV$^2$/fm with $\pm0.3$ GeV$^2$/fm uncertainty.
Due to the rather large statistical uncertainty in our $\hat{q}$ extraction, we cannot make any statements regarding the energy-dependence of $\hat{q}$ at this time. Our results do suggest a range of values for $\hat{q}$ somewhere between 1--1.5 GeV$^2$/fm for the RHIC system.

In future studies, we will extend our work to the temperature dependence of the transport coefficient
and energy-loss as well as to the heavy-quark energy dependence of these quantities. The ultimate
goal of course will be the application of the PCM to heavy quark observables in
ultra-relativistic heavy-ion collisions at the LHC.

\section{Summary}
The present work aims to validate the applicability of the Parton Cascade Model
for the description  of heavy quark evolution in a partonic medium. We have calculated collisional
and radiative energy loss of heavy quarks in an infinite medium at fixed temperature and find good
agreement between the PCM  and analytical calculations for elastic energy loss, but some discrepancies regarding radiative energy loss, which can be understood in terms of the approximations made in the analytical calculations. This is a first important step towards applying
the PCM to the production and evolution of heavy-quarks in a QGP, as produced in collisions of
ultra-relativistic heavy-ions at RHIC and LHC.

\section*{Acknowledgments}
One of us (MY) would like to thank the Nuclear Theory group at Duke University for their hospitality. MY
and DKS have been supported by the DAE, Govt. of India and SAB and CCS acknowledge
support by U.S. department of Energy under grants DE-FG02-05ER41367 and DE-SC0005396.
We are grateful for many
helpful discussions with Berndt M\"uller and Guangyou Qin.








\end{document}